\documentclass{emulateapj}
\usepackage{graphicx}
\usepackage{epstopdf}
\usepackage{mathrsfs}
\usepackage{amsmath}
\usepackage{xspace}
\usepackage{natbib}
\newcommand{\mbh}{\ensuremath{M_{\mathrm{BH}}}\xspace}

\newcommand{\msun}{\ensuremath{M_\odot}\xspace}
\newcommand{\kms}{\ensuremath{\mathrm{km s}^{-1}}\xspace}
\newcommand{\gammavalue}{\ensuremath{0.05_{-0.60}^{+0.29}}\xspace}
\newcommand{\betavalue}{\ensuremath{0.01_{-0.34}^{+0.35}}\xspace}
\newcommand{\mbhvalue}{\ensuremath{5.76_{-1.26}^{+1.76}\times10^6}\xspace}
\newcommand{\rvalue}{\ensuremath{8.92_{-0.55}^{+0.58}}\xspace}
\newcommand{\mbhvaluecombo}{\ensuremath{4.62_{-0.48}^{+0.54}\times10^6}\xspace}
\newcommand{\rvaluecombo}{\ensuremath{8.46_{-0.38}^{+0.42}}\xspace}

\begin{document}
\title{3D stellar kinematics at the Galactic center: measuring the nuclear star cluster spatial density profile, black hole mass, and distance}
\author{T. Do\altaffilmark{1,2,8}, G. D. Martinez\altaffilmark{2,3}, S. Yelda\altaffilmark{4}, A. Ghez\altaffilmark{4}, J. Bullock\altaffilmark{2}, M. Kaplinghat\altaffilmark{2}, J. R. Lu\altaffilmark{5}, A. G. H. Peter\altaffilmark{2,6,7}, K. Phifer\altaffilmark{4}}

\altaffiltext{1}{Dunlap Institute for Astronomy and Astrophysics,
University of Toronto, 50 St. George Street, Toronto M5S 3H4, ON, Canada}
\altaffiltext{2}{Physics and Astronomy Department, University of California, Irvine, CA 92697}
\altaffiltext{3}{The Oskar Klein Center, Department of Physics, Stockholm University, Albanova, SE-10691 Stockholm, Sweden}
\altaffiltext{4}{Physics and Astronomy Department, University of California,
    Los Angeles, CA 90095}
\altaffiltext{5}{Institute for Astronomy, University of Hawaii, HI}
\altaffiltext{6}{Center for Cosmology and Astro-Particle Physics and Department of Physics, Ohio State University, 191 W. Woodruff Ave., Columbus, OH 43210}
\altaffiltext{7}{Department of Astronomy, Ohio State University, 140 W. 18th Ave, Columbus, OH 43210}
\begin{abstract}
\altaffiltext{8}{Dunlap Fellow}
We present 3D kinematic observations of stars within the central 0.5 pc of the Milky Way nuclear star cluster using adaptive optics imaging and spectroscopy from the Keck telescopes. Recent observations have shown that the cluster has a shallower surface density profile than expected for a dynamically relaxed cusp, leading to important implications for its formation and evolution. However, the true three dimensional profile of the cluster is unknown due to the difficulty in de-projecting the stellar number counts. Here, we use spherical Jeans modeling of individual proper motions and radial velocities to constrain for the first time, the de-projected spatial density profile, cluster velocity anisotropy, black hole mass ($M_\mathrm{BH}$), and distance to the Galactic center ($R_0$) simultaneously. We find that the inner stellar density profile of the late-type stars, $\rho(r)\propto r^{-\gamma}$ to have a power law slope $\gamma=0.05_{-0.60}^{+0.29}$, much more shallow than the frequently assumed Bahcall $\&$ Wolf slope of $\gamma=7/4$. The measured slope will significantly affect dynamical predictions involving the cluster, such as the dynamical friction time scale. The cluster core must be larger than 0.5 pc, which disfavors some scenarios for its origin. Our measurement of $M_\mathrm{BH}=5.76_{-1.26}^{+1.76}\times10^6$ $M_\odot$ and $R_0=8.92_{-0.55}^{+0.58}$ kpc is consistent with that derived from stellar orbits within 1$\arcsec$ of Sgr A*. When combined with the orbit of S0-2, the uncertainty on $R_0$ is reduced by 30\% ($8.46_{-0.38}^{+0.42}$ kpc). We suggest that the MW NSC can be used in the future in combination with stellar orbits to  significantly improve constraints on $R_0$. 
\end{abstract}

\section{Introduction}

Due to its proximity, the nuclear star cluster (NSC) at the center of the Milky Way (MW) is the only galactic nucleus for which we are currently capable of resolving individual stars and measure both their proper motion and line-of-sight velocities. This provides us with the unique opportunity to study the dynamical interactions of a star cluster with a supermassive black hole (BH) in unprecedented detail. 

One of the predictions for a dynamically relaxed star cluster with a massive black hole is that there should be a steep increase in stellar density toward the black hole. Sometimes termed the Bahcall and Wolf (BW) cusp, these clusters are predicted to have a power law density profile $\rho(r)\propto r^{-\gamma}$, where $r$ is the physical distance from the black hole, with $\gamma=7/4$ to $3/2$ depending on the relative masses of stars in the cluster \citep{1976ApJ...209..214B,1977ApJ...216..883B}. The red giants in the MW NSC, which constitute the majority of the observable stars in the cluster, are likely old enough (1-10 Gyr) to have formed such a cusp. However, star counts using adaptive optics (AO) spectroscopy and medium-band imaging have shown that the red giants have a very flat projected surface density profile close to Sgr A*, the central black hole \citep{2009A&A...499..483B,2009ApJ...703.1323D,2013ApJ...764..154D}. Due to the effect of projection, it is difficult with number counts alone to constrain the three-dimensional spatial density profile. The flat projected surface density profile even allows for decreasing stellar density toward the black hole, or a 'hole' in the stellar distribution. It is important to measure the spatial density profile of the MW NSC as it may lead us to understand better its formation and evolution \citep[e.g.][]{2010ApJ...718..739M,2012ApJ...750..111A}. The density profile is also important for dynamical considerations, such as the growth of black holes, the effect of dynamical friction, and the predictions for gravitational waves due to the in-spiral of compact objects. 

There has also been much interest over the years in using stellar dynamical measurements of the MW NSC  to constrain the existence of a supermassive black hole at the Galactic center (GC) and measure its mass (M$_{BH}$). A number of approaches at measuring M$_{BH}$ were made using stellar radial velocities and proper motions \citep[e.g.][]{1996ApJ...472..153G,1998ApJ...509..678G,2000MNRAS.317..348G,2001AJ....122..232C} in combination with dynamical modeling. Using an assumed distance to the GC (usually 8 kpc), early measurements of M$_{BH}$ have been made using isotropic \citep{1996ApJ...472..153G} and anisotropic Jeans models \citep{2000MNRAS.317..348G}, mass estimators \citep{1998ApJ...509..678G}, and non-parametric isotropic mass modeling \citep{2001AJ....122..232C}. The measurements for M$_{BH}$ using these methods range from 1.8 to 3.6$\times10^6$ M$_\odot$. However, subsequent measurements using the orbit of S0-2, a star with a semi-major axis of $0.124\arcsec$, have found a higher M$_{BH}$ of $4.1\pm0.6\times10^6$ M$_\odot$ and $R_0 = 8.0\pm0.6$ kpc \citep{2008ApJ...689.1044G,2009ApJ...692.1075G}. Later estimates of \mbh using Jeans modeling of proper motions \citep{2009A&A...502...91S} and in combination with radial velocities \citep{2008A&A...492..419T} has come into closer agreement with the stellar orbits. One source of the discrepancy between these cluster studies may lie in the assumption of the spatial profile of the stars; the earlier works assumed a BW-like cusp, while the later studies assumed a more shallow stellar density profile. However, no studies have so far measured the volume density profile of the MW NSC simultaneously with M$_{BH}$ and $R_0$ to verify consistency with the results from S0-2. 

We present a spherical Jeans model of the inner 0.5 pc of the MW NSC which for the first time, measures the stellar density profile, velocity anisotropy, M$_{BH}$, and $R_0$ simultaneously using individual three-dimensional velocity measurements. In Section \ref{sec:observations} and \ref{sec:sample} we present the observations and sample selection. We discuss the components of our dynamical model in Section \ref{sec:jeans_modeling}, while in Section \ref{sec:results}, we present significant constraints on the cluster parameters as well as M$_{BH}$ and $R_0$. Section \ref{sec:discussion} discusses the implications of these measurements for cusp formation and dynamical calculations involving the cluster. 

\section{Observations}
\label{sec:observations}
Observations of the central 0.5 pc were made using the integral-field spectrograph OSIRIS and the imager NIRC2 on the Keck 2 telescope between 2006 and 2010. These instruments are behind a laser-guide star adaptive optics system (LGS AO) system. The spectroscopic observations are made along the inclination angle of the young stellar disk, at a position angle of 105 deg \citep{2009ApJ...690.1463L} extending out to 14$\arcsec$ from the Galactic center. For more details about the spectroscopic fields and data reduction, see \citet{2013ApJ...764..154D}, where the stellar identifications are reported. Observations and data reduction for the imaging observations are detailed in Yelda et al. (2013, submitted).

\section{Sample Selection and Velocity Measurements}
\label{sec:sample}
In the current study, we only include stars which are identified as late-type in Table 2 from \citet{2013ApJ...764..154D}, as we are interested in the properties of the older component of the NSC. It is important not to include the dynamics from the young stellar population in this region, because at least half of them are distributed in a distinct stellar disk, which would bias the results of spherical Jeans modeling.  In contrast, the old component is likely to be spherically symmetric \citep{2008A&A...492..419T,2010A&A...511A..18S}. In addition, this sample include only stars with all three components of velocity measured (thus limited by the coverage of the OSIRIS observations).  The line-of-sight measurements where obtained by cross-correlation of the late-type stars with an M3II stellar template (HD40239) from the SPEX instrument on IRTF \citep{2009ApJS..185..289R}. The radial velocities are then corrected for the solar motion with respect to the standard of rest\footnote{The velocity correction is performed using the \textit{rvcorrect} task in IRAF. This correction uses a velocity of 20 \kms for the solar motion with respect to the local standard of rest in the direction $\alpha = 18^{h}, \delta = +30\deg$ for epoch 1900 \citep{1986MNRAS.221.1023K}, corresponding to $(u,v,w) = (10, 15.4, 7.8)$ \kms.}. Radial velocity errors are obtained by splitting the spectra for each star into three subsets and measuring the standard deviation of the radial velocities of the three subsets. We are able to measure radial velocities up to $\sim3$ \kms precision, where systematic uncertainties, such as the wavelength solution, then dominate.  The proper motion measurements are made using the data set and reference frame defined by Yelda et al. (2013, submitted). We further only include stars with velocity error less than 100 \kms in any velocity component. The median radial velocity error in our sample is 16 \kms. The median error in $v_x$ is 0.09 mas yr$^{-1}$ and $v_y$ is 0.11 mas yr$^{-1}$ (3.6 \& 4.4 \kms at 8 kpc). The final sample of 265 stars and their velocities are listed in Table \ref{tab:velocities}. 

\begin{deluxetable*}{cccccccccccc}
\tablecolumns{12}
\tablecaption{Measured Stellar Positions and Velocities}
\tablehead{\colhead{Name} & \colhead{K$^\prime$} & \colhead{$\Delta$RA} & \colhead{$\Delta$Dec} & \colhead{$v_x$} & \colhead{$\sigma_{v_x}$} & \colhead{$v_y$} & \colhead{$\sigma_{v_y}$} & \colhead{Epoch$_{xy}$} & \colhead{$v_z$} & \colhead{$\sigma_{v_z}$\tablenotemark{a}} & \colhead{Epoch$_{v_z}$} \\
\colhead{} & \colhead{} & \colhead{(\arcsec)} & \colhead{(\arcsec)} & \colhead{(mas yr$^{-1}$)} & \colhead{(mas yr$^{-1}$)} & \colhead{(mas yr$^{-1}$)} & \colhead{(mas yr$^{-1}$)} & \colhead{} & \colhead{(km s$^{-1}$)} & \colhead{(km s$^{-1}$)} & \colhead{}}
\startdata
S0-17 & 15.9 &  0.0410 & -0.0014 & 7.09 & 0.08 & 24.18 & 0.09 & 2008.177 &  597 &   16 & 2005.500 \\
S0-6 & 14.1 &  0.0296 & -0.3604 & -5.12 & 0.02 & 3.57 & 0.03 & 2007.675 &   87 &    1 & 2008.370 \\
S0-18 & 15.1 & -0.1153 & -0.4173 & -2.66 & 0.04 & 1.82 & 0.04 & 2007.321 & -262 &   19 & 2008.370 \\
S0-27 & 15.6 &  0.1491 &  0.5477 & 0.76 & 0.03 & 3.61 & 0.05 & 2008.449 & -126 &   43 & 2008.560 \\
S0-28 & 15.7 & -0.1411 & -0.4956 & 7.61 & 0.04 & 13.60 & 0.05 & 2007.228 & -329 &   73 & 2008.560 \\
S0-12 & 14.3 & -0.5560 &  0.4124 & 0.98 & 0.02 & 3.88 & 0.03 & 2007.779 &  -29 &    6 & 2008.370 \\
S0-13 & 13.3 &  0.5503 & -0.4119 & 1.80 & 0.02 & 3.64 & 0.03 & 2007.642 &  -38 &    3 & 2006.490 \\
S1-5 & 12.7 &  0.3309 & -0.8978 & -3.84 & 0.02 & 4.71 & 0.03 & 2006.293 &   21 &    2 & 2009.340 \\
S1-10 & 14.7 & -1.1108 & -0.0128 & 4.28 & 0.02 & 2.24 & 0.03 & 2007.804 &  -35 &   17 & 2006.460 \\
S1-31 & 15.7 & -0.9853 &  0.5548 & -0.87 & 0.03 & -2.01 & 0.04 & 2008.473 &  159 &   50 & 2008.370 \\
\enddata
\tablenotetext{a}{Listed are statistical uncertainties. There is $\sim3$ km s$^{-1}$ systematic uncertainty from the wavelength solution.}
\tablenotetext{b}{A full table is published electronically.}
\label{tab:velocities}
\end{deluxetable*}

\section{Jeans modeling}
\label{sec:jeans_modeling}
In order to relate the measured stellar positions and velocities to the cluster properties and the gravitational potential, we will use the spherically symmetric Jeans equation, allowing for velocity anisotropy:
\begin{equation}
\frac{d(\rho_{\star}\sigma_r^2)}{dr}+2\frac{\beta(r)\rho_{\star}\sigma_r^2}{r}=-G\rho_{\star}\frac{M(r)}{r^2},
\end{equation}
where $\sigma_r$ is the dispersion in the radial direction in spherical coordinates. The velocity anisotropy, $\beta$ is defined as in \citet{2008gady.book.....B}:
\begin{equation}
\beta \equiv1-\frac{\sigma_\theta^2+\sigma_\phi^2}{2\sigma_r^2},
\end{equation}
where $\sigma_\theta$, $\sigma_\phi$, and $\sigma_r$ are the velocity dispersion in spherical coordinates. The radial dependence of the anisotropy, $\beta(r)$ is parameterized as:
\begin{equation}
\beta(r)=\frac{\beta_0+\beta_{\infty}(r/r_{\beta})^{\eta}}{1+(r/r_{\beta})^{\eta}},
\end{equation}
where $\beta_o$ is the inner anisotropy, $\beta_\infty$ is the outer anisotropy, $r_\beta$ is the location of the transition, and $\eta$ is the sharpness of the transition. The stellar spatial density profile, $\rho_\star(r)$ is defined to be a broken power law \citep{1995AJ....110.2622L}:
\begin{equation}
\rho_{\star}(r) \propto \left(\frac{r}{r_b}\right)^{-\gamma}\left(1+(r/r_b)^{\delta}\right)^{(\gamma-\alpha)/\delta},
\end{equation}
where $\gamma$ is the inner power law slope, $\alpha$ is the outer slope, $r_b$ is the break radius, and $\delta$ is the sharpness of the transition between the two slopes. The mass profile $M(r)$ is defined to be a point source\footnote{The stellar kinematics are dominated by the gravitational influence of the black hole in the observed region (r$_{infl}=2$ pc). } with the mass of the black hole:
\begin{equation}
M(r) = M_{BH}.
\end{equation}
We also include the distance to the black hole, $R_0$, as a free parameter. The mean velocity of the cluster with respect to the center of the reference frame (Yelda et al. 2013, submitted) is included ($\overline v_{x}$, $\overline v_{y}$, $\overline{v}_{z}$). The total set of model parameters are:
\begin{equation}
\label{eqn:model}
\footnotesize
\mathscr{M}=\{\overline v_{x},\overline v_{y},\overline{v}_{z},r_b,\alpha,\delta,\gamma,r_{\beta},\beta_o,\eta,\beta_{\infty},M_{BH},R_o\}
\end{equation}

Given this set of model parameters and the Jeans equation, we can compute the radial dispersion and project it on the sky to compare to our observed dispersion measurements and its covariances. In order to avoid having to bin the data either radially or by velocity and thus losing information, we choose to compute the likelihoods for each source having an \textit{observed} (projected) velocity vector $\mathbf{V}=\{V_R, V_T, V_z \}$ and a projected distance from Sgr A* of $R$, given the set of model parameters $\mathscr{M}$. Following the methodology in \citet{2011ApJ...738...55M} and \citet{2012JPhCS.372a2016D}, the probability density function (PDF) for each individual star is defined as:

\begin{equation}
\footnotesize
\mathcal{P}(\mathbf{V},R|\mathscr{M})\propto\frac{1}{\sqrt{\left|\mathbf{C}(R)\right|}}\exp\left[-\frac{(\mathbf{V}-\overline{\mathbf{V}})^T C(R)^{-1} (\mathbf{V}-\overline{\mathbf{V}})}{2}\right].
\end{equation}
The combined likelihood of the whole sample is the product of the PDF for each star, $i$:
\begin{equation}
\mathcal{L}(\mathscr{M})=\prod_i\mathcal{P}(\mathbf{V_i},R_i|\mathscr{M})
\end{equation}
The likelihood is proportional to $R_0^2$, the distance to the Galactic center through the conversion between angular velocity into physical units. 

The covariance matrix ($\mathbf{C}$) of the \textit{intrinsic} moments of the velocity components ($\mathbf{v}=\{v_R,v_T,v_z\}$ in cylindrical coordinates):

\begin{align}
\mathbf{C}\equiv\frac{1}{\Sigma(R)}\int_{-\infty}^{\infty}{\rho_\star<\mathbf{v}^T\mathbf{v}>dz}
\end{align}
Where $\Sigma(R)$ is the projected surface density profile. Equivalently this covariance matrix came be expressed as a function of the anisotropy $\beta(r)$:
\begin{align}
\label{eqn:covariance}
\mathbf{C}=&\frac{2}{\Sigma(R)}\int_R^{\infty}{\mathbf{A}\frac{r\rho_{\star}\sigma_r^2(r)}{\sqrt{r^2-R^2}}dr},\\
\mathbf{A}=&\left(\begin{array}{ccc}
\left[1-\beta\left(1-\frac{R^2}{r^2}\right)\right]&0&\left[\beta\frac{R}{r}\sqrt{1-\frac{R^2}{r^2}}\right]\\
0&\left[1-\beta\right]&0\\
\left[\beta\frac{R}{r}\sqrt{1-\frac{R^2}{r^2}}\right]&0&\left[1-\beta\frac{R^2}{r^2}\right]\end{array}\right).
\end{align}
Some of the off-diagonal terms are zero because of our assumption of spherical symmetry. The error matrix to account for measurement error is (the errors are assumed to be normally distributed and uncorrelated):
\begin{align}
\mathbf{\mathcal{E}}_{RT}\equiv\left(\begin{array}{ccc}
\epsilon_{V_R}^2&0&0\\
0&\epsilon_{V_{T}}^2&0\\
0&0&\epsilon_{V_z}^2\end{array}\right) 
\end{align}
The total projected covariance matrix is therefore:
\begin{equation}
\mathbf{C}_{RT,total}=\mathbf{C}_{RT}+\mathbf{\mathcal{E}}_{RT}
\end{equation}
If Cartesian coordinates are chosen, such that $\mathbf{v} = (v_x,v_y,v_z)$ then $\mathbf{A}$ in Equation \ref{eqn:covariance} is:
\begin{align}
\mathbf{A}_{xy}=&\left(\begin{array}{ccc}
\left[1-\beta\left(1-\frac{x^2}{r^2}\right)\right]&\left[\beta\frac{xy}{r^2}\right]&\left[\beta\frac{x}{r}\sqrt{1-\frac{R^2}{r^2}}\right] \\
\left[\beta\frac{xy}{r^2}\right]&\left[1-\beta\left(1-\frac{y^2}{r^2}\right)\right]&\left[\beta\frac{y}{r}\sqrt{1-\frac{R^2}{r^2}}\right] \\
\left[\beta\frac{x}{r}\sqrt{1-\frac{R^2}{r^2}}\right]&\left[\beta\frac{y}{r}\sqrt{1-\frac{R^2}{r^2}}\right]&\left[1-\beta\frac{R^2}{r^2}\right]
\end{array}\right).
\end{align}

We take a Bayesian approach along with the Multi-Nest sampling algorithm \citep{2009MNRAS.398.1601F} to compute the posterior PDFs with linear flat priors for all parameters (Table \ref{tab:priors}).  

\section{Results}
\label{sec:results}
\begin{deluxetable*}{lccccccc}[ht]
\tablecolumns{8}
\tablecaption{Model Parameters, Priors, and Fits}
\tablehead{\colhead{Parameter} & \colhead{Description} & \colhead{Unit} & \colhead{Lower Limit} & \colhead{Upper Limit} & \colhead{Fit 1\tablenotemark{a}} & \colhead{Fit 2\tablenotemark{b}} & \colhead{Fit 3\tablenotemark{c}} }
\startdata
$\overline{v}_x$ & central velocity in x direction & km s$^{-1}$ & -100 & 100  & $24.99_{-7.74}^{+7.96}$  & $21.99_{-7.82}^{+7.87}$  & $23.13_{-7.47}^{+7.60}$ \\
$\overline{v}_y$ & central velocity in y direction & km s$^{-1}$ & -100 & 100  & $10.06_{-8.18}^{+8.05}$  & $8.11_{-7.78}^{+7.80}$  & $9.40_{-7.98}^{+7.69}$ \\
$\overline{v}_z$ & central velocity in z direction & km s$^{-1}$ & -100 & 100  & $-6.11_{-7.32}^{+7.49}$  & $-7.23_{-8.00}^{+8.07}$  & $-6.41_{-7.39}^{+7.43}$ \\
$r_\beta$ & break radius in $\beta(r)$ & pc & 0 & 2  & $0.93_{-0.28}^{+0.54}$  & \nodata & $1.13_{-0.53}^{+0.54}$ \\
$\beta_o$ & inner anisotropy   &  &-3 & 1  & $0.01_{-0.34}^{+0.35}$  & $0.00$  & $-0.13_{-0.30}^{+0.39}$ \\
$\eta$ & sharpness in $\beta(r)$ transition & & 0 & 10  & $4.55_{-2.72}^{+3.46}$  & \nodata & $4.24_{-2.80}^{+3.81}$ \\
$\beta_\infty$ & outer anisotropy & & -5 & 1  & $-2.72_{-1.43}^{+1.50}$  & $0.00$  & $-2.01_{-1.81}^{+1.25}$ \\
$r_b$ & break radius for $\rho_\star$ & pc & 0 & 2  & $1.56_{-0.25}^{+0.26}$  & $1.53_{-0.25}^{+0.28}$  & $1.51_{-0.24}^{+0.27}$ \\
$\gamma$ & inner slope of $\rho_\star$ & & -5 & 2  & $0.05_{-0.60}^{+0.29}$  & $0.16_{-0.30}^{+0.25}$  & $0.22_{-0.30}^{+0.22}$ \\
$\delta$ &  sharpness in $\rho_\star$ transition & &  0 & 10  & $6.87_{-2.65}^{+2.10}$  & $6.83_{-2.55}^{+2.08}$  & $6.81_{-2.62}^{+2.16}$ \\
$\alpha$ & outer slope in $\rho_\star$ & &  3 & 10  & $5.94_{-2.11}^{+2.68}$  & $5.88_{-2.10}^{+2.68}$  & $6.31_{-2.19}^{+2.47}$ \\
$M_{BH}$ & black hole mass & $\times10^6$ M$_\odot$ & 3 & 8  & $5.76_{-1.26}^{+1.76}$  & $3.77_{-0.52}^{+0.62}$  & $4.62_{-0.48}^{+0.54}$ \\
$R_{o}$ & distance to GC & kpc & 5 & 10  & $8.92_{-0.55}^{+0.58}$  & $8.12_{-0.41}^{+0.43}$  & $8.46_{-0.38}^{+0.42}$ \\
\enddata
\tablenotetext{a}{Fit using MW NSC stars.}
\tablenotetext{b}{Fit with isotropic velocity distribution.}
\tablenotetext{c}{Fit 1 with $R_o$ and M$_{BH}$ priors from the orbit of S0-2 \citep{2008ApJ...689.1044G}}.
\label{tab:priors}
\end{deluxetable*}

\begin{figure*}
\center
\includegraphics[width=6.5in]{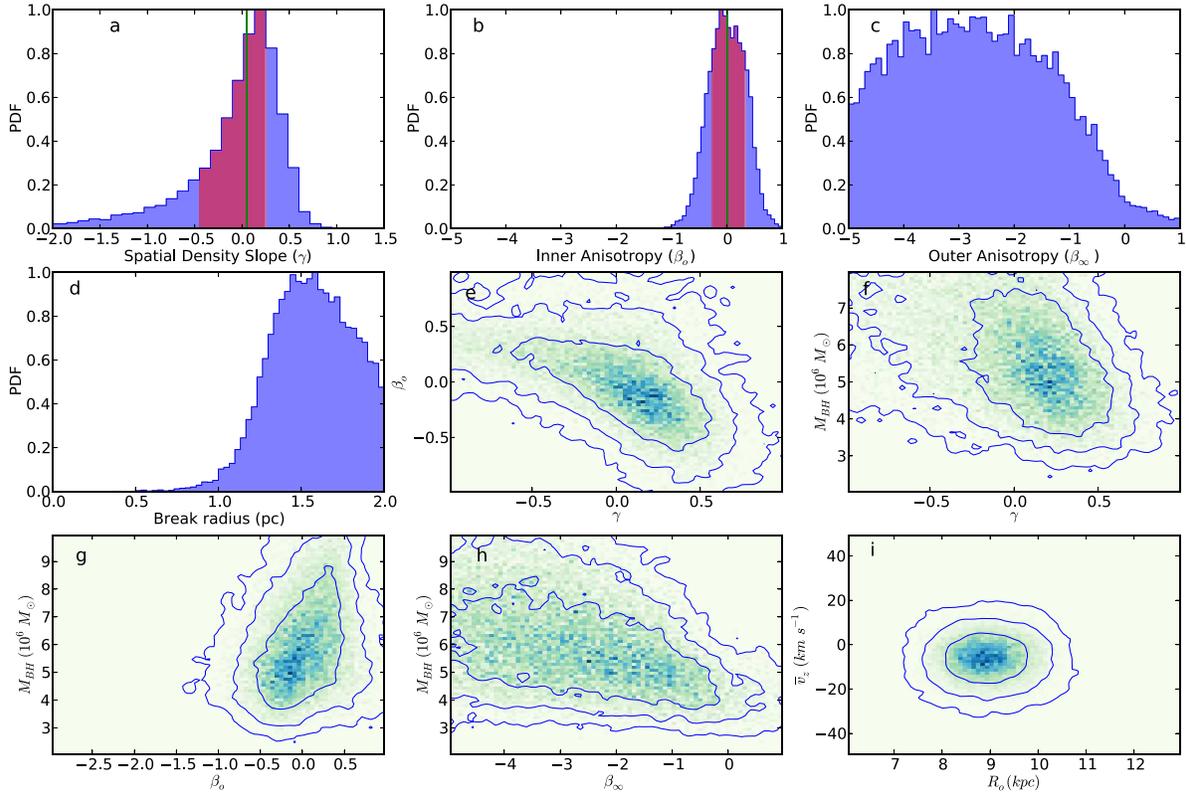}
\caption{The first row of plots shows the posterior distributions for the cluster parameters for the case where both \mbh and $R_0$ are allowed to be free. The measurements have 68\% central confidence intervals of: (a) $\gamma=\gammavalue$, (b) $\beta_o=\betavalue$. The constraints on $\gamma$ and $\beta_o$ are robust, but the core radius (c) and outer anisotropy $\beta_\infty$ (d) are dominated by the priors because the data is limited to $R < 0.5$ pc. The joint posteriors between (e) $\gamma$ and $\beta_0$, and (f) $\gamma$ and \mbh show strong correlation. The third row shows the joint posteriors between (g) $\beta_0$ and \mbh, (h) $\beta_\infty$ and \mbh, and (i) the distance to the Galactic center, $R_0$ and the bulk cluster velocity along the line of sight, $\overline v_z$. There is no correlation between $\overline v_z$ and $R_o$.}
\label{fig:posteriors}
\end{figure*}

We find the combination of kinematic data and Jeans modeling is able to significantly measure most of the cluster parameters, including $\gamma$, $\beta_o$, M$_{BH}$, and $R_0$.  For the case where M$_{BH}$ and $R_0$ are allowed to be free, we find the 68\% central confidence interval for  $\gamma=\gammavalue$, $\beta_o=\betavalue$, $\mbh=\mbhvalue$ \msun, and $R_0=\rvalue$ kpc (Figure \ref{fig:posteriors}). We find that there is a strong correlation between $\beta_o$ and $\gamma$, as well as between $\gamma$ and \mbh (Figure \ref{fig:posteriors}). The fits are insensitive to the mean velocity offset in each direction ($\overline v_x, \overline v_y, \overline v_z$), which are not correlated with any of the cluster parameters. As an example, we show the joint PDF of $v_z$ and $R_0$ in Fig. \ref{fig:posteriors}.

\begin{figure*}
\center
\includegraphics[width=6.5in]{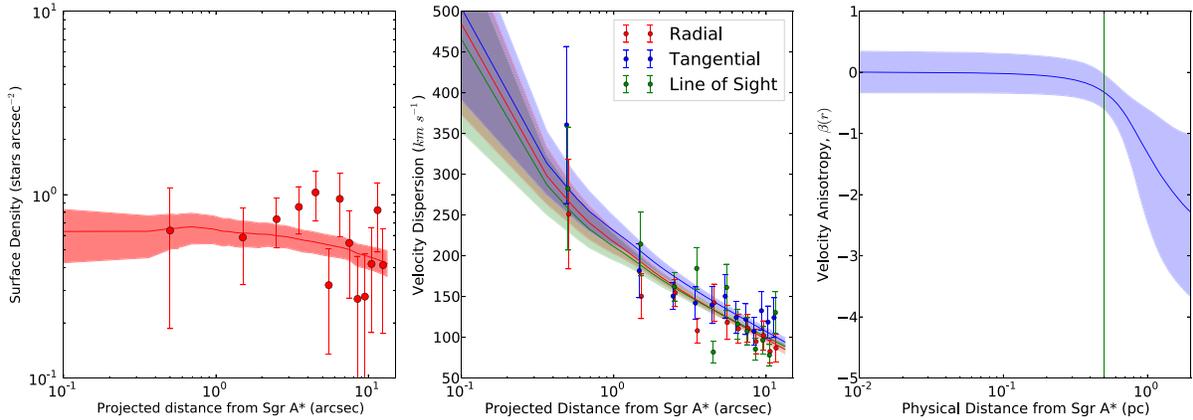}
\caption{Comparison between the results of the Jeans modeling with observations. The observations are binned purely for illustrative purposes; the model uses individual likelihoods of the position and velocities of the stars to constrain cluster parameters. \textbf{Left:} The predicted projected surface density profile with the 1 $\sigma$ shaded region and the observed (not completeness corrected) number density from \citet{2013ApJ...764..154D}. \textbf{Center:} The predicted projected velocity dispersion as a function of projected distance from Sgr A* in the three velocity components and the 1 $\sigma$ deviation from the most probable fit. \textbf{Right:} the central 68\% confidence interval for velocity anisotropy as a function of physical radius from the black hole. The velocity anisotropy is flat to the edge of our data (green line); beyond this region, the constraints are prior dominated.}
\label{fig:results}
\end{figure*}

In order to test the robustness of these results, we also re-analyze the data with variations of assumptions about the cluster parameters: (1) using the joint PDF of $M_{BH}$ and $R_o$ from the orbit of S0-2 \citep[$M_{BH}=4.1\pm0.6\times10^6$ M$_\odot$, $R_0=8.0\pm0.6$ kpc]{2008ApJ...689.1044G}, (2) allowing for extended mass ($M(r)=M_{BH}+M_{stars}(r)$), (3) allowing for cluster rotation. We find that the inner slope value is very robust to changes in model assumptions, varying only within the range of statistical uncertainties. Adding the priors on M$_{BH}$ and $R_0$, shifts $\gamma$ by less than 1 $\sigma$,  with the uncertainties staying nearly the same. We find no significant constraints on the mass from stars in this region. The present data show no significant evidence for rotation, even when it is allowed in the fit. Within the radial range of the present dataset (0.5 pc), the velocity dispersion dominates over the large scale rotation that is seen at larger scales of 1 to 2 pc \citep{2008A&A...492..419T,2009A&A...502...91S}. We tabulate the resulting central confidence intervals for the cluster-only fit as well as for the addition of the constraints on M$_{BH}$ and $R_0$ from S0-2 in Table \ref{tab:priors}.

We also examine the effect of including velocity anisotropy by doing an isotropic fit. We find that the inner slope becomes steeper with $\gamma = 0.16^{+0.25}_{-0.30}$, but is consistent within 1 $\sigma$. The inferred black hole mass and $R_0$ also decreases, along their statistical uncertainties (Table \ref{tab:priors}).  We note that inferred $R_0$ has a direct impact on the inferred anisotropy and vice versa.

It has been argued that the non-negativity of the distribution function imposes the constraint $\gamma \geq \beta_0 +1/2$ \citep{2006ApJ...642..752A}. This relation is violated in large parts of the $\beta_0-\gamma$ preferred region (Figure \ref{fig:posteriors}e) and this issue deserves a separate investigation. Including this limit will likely result in slightly steeper $\gamma$ and increased tangential anisotropy. A distribution function analysis similar to that of \citet{2006ApJ...643..210W} will be useful to confirm the present results.

We are unable to place strong constraints on the size of the core profile using the current dataset, except that it must lie beyond 0.5 pc. This lower-limit is consistent with surface number density profiles from \citet{2007A&A...469..125S} show that the cluster core should lie between about 0.4 to 0.6 pc, where the profile turns over to the form of $1/R^{0.8}$. 

Similarly, constraints on the velocity anisotropy is best in the region that the data samples. The posterior distribution for the inner anisotropy ($\beta_0$) is well constrained while the outer anisotropy ($\beta_\infty$) is prior dominated (Fig. \ref{fig:posteriors}).

The Jeans model shows definitively that the Galactic center has only a very shallow cusp of red giants, and that it is inconsistent with the predictions of dynamical relaxation of $\gamma = 3/2$ to $7/4$ \citep{1977ApJ...216..883B}. Number counts from the surface density profile had previously placed upper-limits on the slope of the spatial density profile of $\gamma < 1.0$ at 99.7\% confidence, which allowed for a complete lack of late-type giants close to Sgr A* (or a `hole' in the stellar distribution). The addition of 3D kinematics shows that there must be late-type stars near the black hole, though the stellar density may decrease toward the black hole. 

\begin{figure*}
\center
\includegraphics[width=6.6in]{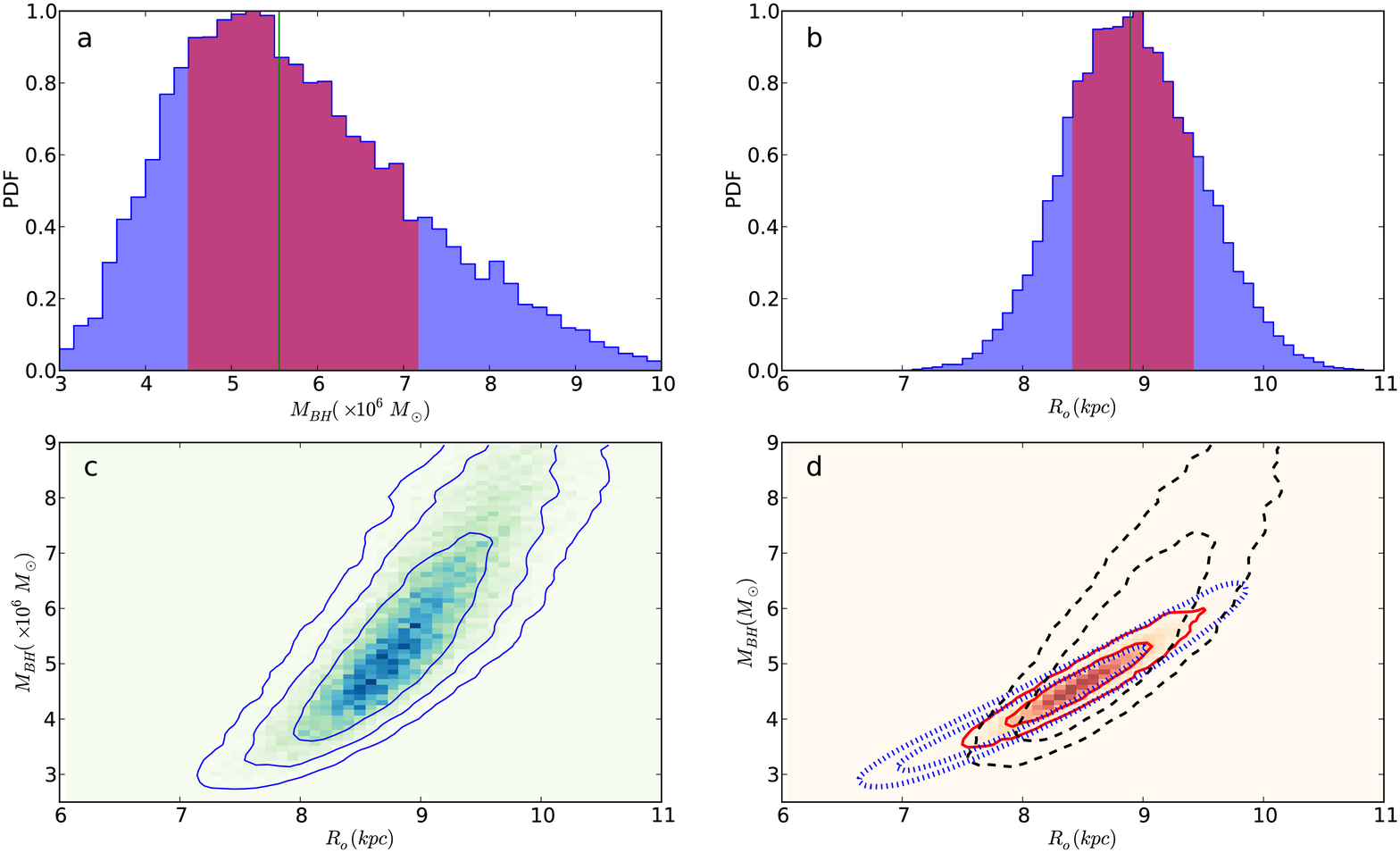}
\caption{The PDFs from the Jeans modeling analysis of the NSC for (a) \mbh and (b) $R_0$, marginalized over all other parameters. The central 68\% confidence interval is shown in shaded red ($\mbh = \mbhvalue$ \msun, $R_o= \rvalue$ kpc). Panel (c) shows the joint PDF of M$_{BH}$ and $R_0$, showing the correlation in the two parameters. The contours are at confidence levels of 68\%, 95\%, and 99.73\%. Panel (d) shows the joint PDF after the addition of the constraints from the orbit of S0-2 \citep{2008ApJ...689.1044G}; the PDF from the cluster (dashed black) and S0-2 (dotted blue) are oriented in different directions, so their combination leads to stronger constraints and can improve the measurement of $R_0$ by 30\% compared to using either methods alone ($\mbh = \mbhvaluecombo$ \msun, $R_0=\rvaluecombo$ kpc).}
\label{fig:mbh}
\end{figure*}

\section{Discussion \& Conclusion}
\label{sec:discussion}
The theoretical explanations for the unrelaxed stellar density profile at the Galactic center can be roughly classified as either as slow and secular, or from disruptive events. Secular explanations include: resonant relaxation of the cluster driving stars more efficiently than two-body relaxation into the black hole loss cone \citep{2011ApJ...738...99M}; collisions between the red giants and other stars or stellar remnants \citep[e.g.][]{2009MNRAS.393.1016D}; or tidal stripping of the red giants \citep[][]{2005ApJ...624L..25D,1991ApJ...370...60M}. While these mechanisms may all occur and can produce core-like profiles, they are generally only effective at $r < 0.1$ pc and are not likely able to fully explain a core with $r>0.5$ pc. Disruptive mechanisms such as the infall of a massive black hole \citep[e.g.][]{2006MNRAS.372..174B,2010ApJ...718..739M} or a globular cluster \citep{2012ApJ...750..111A} can produce very large cores depending on the mass of the infalling object.  While these events are very infrequent, the long two-body dynamical relaxation time in this region would require about 1-10 Gyr to regenerate a BW cusp. To test the theory of an infalling black hole will require theoretical predictions for the motion of stars after the passage of the massive object. For example, tangential anisotropy observed may be indicative of the effect of core scouring by a massive black hole, though other mechanisms may also result in this effect. The radial profile of the velocity anisotropy could also potentially offer a constraint on its formation. Simulations of merging globular clusters from \citet{2012ApJ...750..111A} show a linearly decreasing $\beta$ outward from the cluster center at $\beta = 0$. Our dataset shows a generally flat velocity anisotropy profile within the region with data (Fig. \ref{fig:results}). Future kinematic data from outside the central 0.5 pc will be necessary to provide better constraints on these scenarios. 

We also investigate the potential for combining the cluster measurements with that from stellar orbits to reduce statistical uncertainties in M$_{BH}$ and $R_0$. Figure \ref{fig:mbh}a shows the joint probability distribution for M$_{BH}$ and $R_0$ based on the cluster alone, while Figure \ref{fig:mbh}b shows the addition of the PDF on $M_{BH}$, $R_o$, and $v_z$ from the orbit of S0-2 \citep{2008ApJ...689.1044G}. Because the two methods of measuring M$_{BH}$ and $R_0$ have different degeneracies, their combination provides better constraints on both parameters. The joint fit has $\mbh=\mbhvaluecombo$ \msun and $R_0=\rvaluecombo$ kpc. The uncertainty in $R_0$ has improved by about 30\%, from 0.6 kpc using either S0-2 or the cluster alone. This points toward a method for significantly improving the measurement of $R_0$, which is important for scaling other measurements of the MW, such as the Galaxy's total mass and shape \citep[e.g][]{2000MNRAS.311..361O}. However, there may be a number of systematic effects that would need to be quantified before this method can be employed. For example, errors in establishing the reference frame for astrometry can lead to shifts in the orbit determination \citep{2010ApJ...725..331Y}. If the cluster is highly aspherical or triaxial, the cluster measurements can also be biased. These systematic errors can be quantified in the near future with more dynamical data and comparisons with N-body solutions. Kinematic measurements of the MW NSC were among the first methods used for measuring M$_{BH}$ and it now holds renewed promise for significantly improving measurements of both M$_{BH}$ and $R_0$ using existing data.

Given the richness of the dynamical data available for the Galactic center, more sophisticated approaches to dynamically model the cluster can be used in the future. While spherical Jeans modeling appears to be a suitable fit for the inner 0.5 pc of the Galaxy, on larger scales, axisymmetric or triaxial models will be more flexible in incorporating deviations from sphericity and rotation. The phase space distribution of the cluster can also be mapped directly, given that 5 out of 6 of the phase-space parameters have been measured for each star. These kinematic measurements will not only reveal more about the properties and origins of the MW NSC, but also drive the development of better dynamical models of star clusters.  

We thank the anonymous referee for helpful comments. AHGP was partially supported by a Gary McCue Fellowship through the Center for Cosmology at UC Irvine

\end{document}